\documentclass[aps,pre,twocolumn,groupedaddress]{revtex4-1}
\usepackage[latin9]{inputenc}
\usepackage{amsmath}

\makeatletter

\usepackage{graphicx}
\graphicspath{ {figures/} }
\usepackage{epstopdf}
\usepackage{subcaption}
\usepackage{float}
\usepackage{listings}
\makeatother

\begin{document}

\title{Precise spatial spatial memory in local random networks}

\author{Joseph L. Natale}

\affiliation{Department of Physics, Emory University, Atlanta, GA 30322, USA }

\author{H. George E. Hentschel}
\affiliation{Department of Physics, Emory University, Atlanta, GA 30322, USA}

\author{Ilya Nemenman}
\affiliation{Department of Physics, Department of Biology, and Initiative in Theory and Modeling of Living Systems, Emory University, Atlanta, GA 30322, USA}

\date{November 15 2019}

\begin{abstract}
Self-sustained, elevated neuronal activity persisting on time scales of ten seconds or longer is thought to be vital for aspects of working memory, including brain representations of real space. Continuous-attractor neural networks, one of the most well-known modeling frameworks for persistent activity, have been able to model crucial aspects of such spatial memory. These models tend to require highly structured or regular synaptic architectures. In contrast, we elaborate a geometrically-embedded model with a local but otherwise random connectivity profile which, combined with a global regulation of the mean firing rate, produces localized, finely spaced discrete attractors that effectively span a 2D manifold. We demonstrate how the set of attracting states can reliably encode a representation of the spatial locations at which the system receives external input, thereby accomplishing spatial memory via attractor dynamics without synaptic fine-tuning or regular structure. We measure the network's storage capacity and find that the statistics of retrievable positions are also equivalent to a full tiling of the plane, something hitherto achievable only with (approximately) translationally invariant synapses, and which may be of interest in modeling such biological phenomena as visuospatial working memory in two dimensions.
\end{abstract}


\maketitle


\section{Introduction}
\label{intro}

Biological implementations of working memory bridge the gap between two fundamentally disparate time scales: single neurons process information in $\sim 10^{-3}$s, whereas organisms interact with their external environments over durations of $\sim 1$s or longer. For species from fruit flies to primates, this extension of time scales is reflected at the neural level by elevated spiking activity that persists while a particular memory is being accessed~\cite{goldman1995cellular}.

These excitations tend to be highly localized: for various types of working memory tasks across brain regions, firing rates for only a subset of selectively receptive neurons appear to become elevated~\cite{fuster1971neuron,funahashi1989mnemonic,fuster1981inferotemporal,gnadt1988memory}. Traditionally, these units are considered to be responsible for maintaining the memory, and their so-called \emph{persistent activity}, which can last anywhere from tens of seconds to several minutes, is thought to underlie a multitude of well-studied neural computations~\cite{macneil2011fine} (see Ref.~\cite{lundqvist2018working} for an alternative viewpoint). While the mechanistic drivers of persistent activity are not fully understood -- both single-cell and network-level explanations have been proposed over the last several decades, but their relative contributions remain under debate~\cite{zylberberg2017mechanisms} -- attractor neural network models have provided phenomenological descriptions of persistent firing states as fixed points or stable manifolds of the neural dynamics~\cite{seung1996brain,amit1997model,compte2000synaptic}.

Attractor neural networks were first developed within the context of discrete, long-term associative memory, where each attracting state in a multistable system represented a distinct, stored memory~\cite{hopfield1982neural}. Continuous-valued variants have since been able to model transient memories, like the firing activity responsible for maintaining an animal's eye position between saccades in one dimension~\cite{seung1996brain} or its heading direction in a 2D environment~\cite{zhang1996representation}. To be useful in this context, attractor networks must typically incorporate highly structured or precisely tuned connection topologies. For instance, the synaptic connectivity matrices in Ref.~\cite{seung1996brain} satisfy stringent spectral tuning properties that allow certain firing patterns to persist indefinitely. This need for nontrivial structure is quite general: it allows models of persistent activity to ensure the requisite balance between excitation and inhibition, which in turn renders a circuit capable of memory~\cite{zylberberg2017mechanisms}.

Recently, a biological instance of continuous attractor dynamics was traced to a circuit in \emph{Drosophila} that respects one version of these topological constraints~\cite{kim2017ring}. It has been suggested that the fly computation derives from high-level network properties -- topological configuration, local excitation, and long-range inhibition -- rather than ``fine-scale" details like synaptic weights~\cite{kakaria2017ring}. Yet it is not clear that networks with \emph{random} weights, or unstructured connectivities, can perform similar computations. Indeed, random excitatory-inhibitory networks have been shown to be capable of various complex computations, including conjunctive encoding for input classification~\cite{george2018random} and, in the balanced case, emergent selectivity in the context of evidence integration tasks~\cite{sederberg2019randomly}.

In this article, we ask how well a minimally structured, randomly weighted network model can perform a spatial memory task of the kind previously thought~\cite{kriener2014dynamics} to need tuned, regular topologies. To do this, we study the firing-rate dynamics of a system with local but otherwise random connections. The network is spatially extended, and we show that it is able to encode the locations of external stimuli as elevated firing activity in the region near stimulation. In other words, it is capable of spatial memory. We introduce this system in Section~\ref{Models}, and computationally measure its capacity for distinguishing different stimulation locations in Section~\ref{results}. We conclude by discussing how the model relates to previous work, and how it might be extended, in Section~\ref{discussion}. Our intent is not to model any specific biological system, but to demonstrate how computations similar to those of persistent, continuous attractors are theoretically possible in random, balanced networks.

\section{Model and Methods}
\label{Models}

The network $\mathcal{G}=(\lbrace i\rbrace,\lbrace{J_{ij}\rbrace})$ consists of $N$ excitatory rate neurons~\cite{seung1996brain,druckmann2012neuronal},
embedded on a square plane of side length $L$, and connections $\lbrace{J_{ij}\rbrace}$ pointing from neuron $j$ to neighbor $i$ ($i,j=1 \dots N$). Specifically, we choose a set of spatial point coordinates $X=\{(x_1,y_1), ..., (x_N,y_N)\}$, where each pair $\vec{x}_i=(x_i,y_i)$ is an independent random sample from the bivariate uniform distribution on the interval $[0, L]$. This system has uniform spatial density $\sigma=\frac{N}{L^2}$, which is equivalent to $\frac{L}{\sqrt{N}}\equiv\lambda$ as the average inter-neuron separation.

With matrix elements $\lbrace d_{ij} \rbrace$ representing the Euclidean distances between neurons $i$ and $j$, we assign a nonzero value to the synapse strength $J_{ij}$ if $d_{ij}<\xi$, where $\xi \ll L$. We prohibit autapses, or self-loops, and invoke periodic boundary conditions in the calculation of $d_{ij}$. For convenience and uniformity, we present all results using the reference plane $[0,L]\times[0,L]$. In all that follows, $L=1$ and $\xi=0.06L$ unless otherwise specified. We also choose $N=2^{12}$, which fixes $\lambda \approx 0.016 (L) \approx 0.26\xi$.

Choosing a value for $\xi$ which is small relative to $L$ ensures that connections remain short-ranged, and that the resulting network is sparse. We argue later that choosing a set of connections $\lbrace J_{ij} \rbrace$ that is too short or too long-ranged diminishes the ability of the network to support multiple nontrivial memory states. Quantitatively, since each neuron $i$ interacts with $\sim \pi\xi^2\sigma$ downstream neighbors, a typical network realization $\mathcal{G}$ encompasses $\sim \pi N^2\left(\xi/L\right)^2$ synapses, or about $1\%$ of all possible connections.

The connection strengths, or synaptic efficiacies, are
\begin{equation}
J_{ij} = \left\{
        \begin{array}{ll}
           \sim P(\mu,\sigma), & d_{ij} < \xi \text{ and } j\neq i, \\
            0, & d_{ij} \geq \xi \text{ or } j = i,
        \end{array}\right.
\label{eq:J}
\end{equation}
where each $J_{ij}$ is an independent draw from $P(\mu,\sigma)$, representing a lognormal distribution (as argued for in Ref.~\cite{song2005highly} and elsewhere; we explored other distributions, but found no qualitative differences in the results). Since by definition lognormal random variables are positive definite, $J_{ij} > 0$ for all outgoing connections: all neurons are excitatory. In what follows, $\mu=-0.702$ and $\sigma=0.8752$ (by convention, these parameters refer to the associated normal distribution). These values were taken from fits done during experimental investigations of neural circuit properties in the rat visual cortex~\cite{song2005highly}.

As emphasized above, persistent activity typically demands a fine balance between excitation and inhibition, while our connectivities encompass no explicit inhibition.
Therefore, we choose to model inhibition indirectly, imposing its main effect -- which we assume is to stabilize the system's total firing activity to a constant value~\cite{roudi2008representing,monasson2013crosstalk} -- directly. In particular, we insert a term into the usual nonlinear firing-rate equations~\cite{seung1996brain,druckmann2012neuronal} to represent nonlocal inhibitory interactions. In summary, in the absence of synaptic or external inputs, the firing-rate activity $r_i(t)$ decays exponentially over the intrinsic time scale $\tau$. Otherwise, $r_i(t+dt)$ is determined by integrating a nonlinear function of combined input currents $\sum_j J_{ij} r_j(t)$ the from upstream neighbors $j$ and external drive $I_i(t)$ over the short interval $dt \ll \tau$. Thus, for constant $a>0$,
\begin{align}
  \tau\frac{dr_i}{dt} &= -r_i + aN \left(\frac{h_i}{\sum_j h_j}\right), \label{eq:dynamics} \\
  h_i &= f\left(\sum_j J_{ij}r_j+I_i(t)\right). \label{eq:activation}
\end{align}

This system will ultimately approach a steady state for which $\sum_i r_i(t\gg \tau) = aN$: global inhibitory interactions, implemented by the second, ``activation," term in Eq.~(\ref{eq:dynamics}), create the desired balance. This can be verified by solving for the steady-state conditions $\frac{dr_i}{dt}=0$. The parameter $a$ in Eqs.~(\ref{eq:dynamics}-\ref{eq:activation}) can be thought of as the system's baseline firing level (the rate at which all neurons would fire if they were to fire at equal rates in the steady state). A complementary interpretation, related to the fraction of active cells in the steady state, will be addressed in detail later. We set $a=0.02$ and, without loss of generality, choose $\tau=1$ so that time is measured in unit of $\tau$.

Finally, we adopt for the nonlinearity a version of the firing-rate function introduced by Ref.~\cite{zhang1996representation},
\begin{equation}
f(x)=\alpha \cdot \left\lbrace  \ln{ [ 1+\ln{ (1+e^{\beta(x-\gamma) } ) } ] }  \right\rbrace  ^{\delta},
\label{sigmoid}
\end{equation}
with $\alpha=18$, $\beta=0.5$, $\gamma=16$, and $\delta=1.5$. We selected these values to place activations $\lbrace h_i \rbrace$ in a biological range (tens or less, if measured in Hz) for arguments $x>0$ spanning two orders of magnitude, with $f(0)\sim10^{-4}\approx 0$.
The reason for the choice given by Eq.~(\ref{sigmoid}) is that the gain of this curve increases at a value away from zero, and that its behavior in the limit of large inputs is nonsaturating over two orders of magnitude in $x$. These attributes are intended to better approximate the biological reality~\cite{barbieri2008can}, as compared with the sigmoidal thresholding functions commonly used in artificial networks (which tend to feature inflection points near values corresponding to zero net input). We note that both of these properties are also satisfied by the ReLu (Rectified Linear unit) activation function~\cite{hahnloser2001permitted}, also commonly used in machine learning.

For a realization $\mathcal{G}$ with dynamics given by Eqs.~(\ref{eq:dynamics}-\ref{eq:activation}), we would like to quantify how this system performs as a spatial memory architecture. In particular, if a group of neurons local to an arbitrary region of the plane is stimulated externally, can the system sustain a persistent representation of their coordinates? How many distinct stimulation sites can the system reliably encode?

To measure the number of resolvable sites, we perform $n_{\text{trials}}$ ``external stimulation" computational experiments, sequentially, in Matlab. First, we initialize the system, creating a network realization $\mathcal{G}$ by selecting values for the neuron positions $X$ and connection strengths $\lbrace{ J_{ij}\rbrace}$. We then set the firing rates of all neurons $i=1 \dots N$ to $r_i(0)=a$ and evolve Eqs.~(\ref{eq:dynamics}-\ref{eq:activation}) from $t=0$ to $t=100\tau$, well beyond the point at which the individual firing rates stabilize, using the built-in Runge-Kutta~(4,5) solver with $I_i(t)=0$. The result can be a strong excitation, confined to a local region of the plane, or a fully \emph{delocalized} firing state in which all neurons participate with rates near $a$. In either case, the rates do not change in time (this holds even if the system is initialized randomly, with rates that sum to the steady-state value $aN$, instead of uniformly).

To ensure that the system can switch out of this state, we perform a single external stimulation, abitrarily targeting the visual center of the plane, according to the following protocol. With the aformentioned state serving as our initial condition, we locate all neurons contained within an ``input" patch of area $\pi \rho^2$ (for now, we choose $\rho=\xi=0.06L$) centered at $\vec{x}_{\text{stim}}=(0.5,0.5)$. For this subset of system elements only, we set
\begin{equation}
I_i(t) = A \left( 1- \Theta(t-\Delta t) \right) = \left\{
        \begin{array}{ll}
           A, & t < \Delta t, \\
            0, & t \geq \Delta t,
            \label{eq:input}
        \end{array}
        \right.
\end{equation}
where $\Theta(t)$ denotes the Heaviside step function, and $\Delta t = 5\tau$. We again solve Eqs.~(\ref{eq:dynamics}-\ref{eq:activation}), integrating until $T=40\tau$, sufficient time for the network to reach a persistent state.

We then repeat this protocol for $n_{\text{trials}}$ iterations, each time sampling a random position $\vec{x}_{\text{stim}} = (x_{\text{stim}}, y_{\text{stim}})$ from a uniform grid of $10^4$ finely-spaced points superimposed on the plane (that is, separated by $dL = 10^{-2}L$), to serve as the set of stimulation centers. The resulting state $\lbrace r_i(t=T) \rbrace$ then becomes the new initial condition for the following trial, representing re-stimulation and new memory formation. We set $n_{\text{trials}}=k\cdot(L/dL)^{2}$, partitioning stimulations into $k$ successive groups of $(L/dL)^{2}$ trials that are each composed of independent random permutations of the full list of available gridpoints $\lbrace {x_{\text{stim}}} \rbrace$.

\section{Results}
\label{results}

\subsection{Network supports multiple stable attractors}
\label{results__supportsattractors}

\begin{figure}[ht]
  \centering
  \includegraphics[width=1\linewidth]{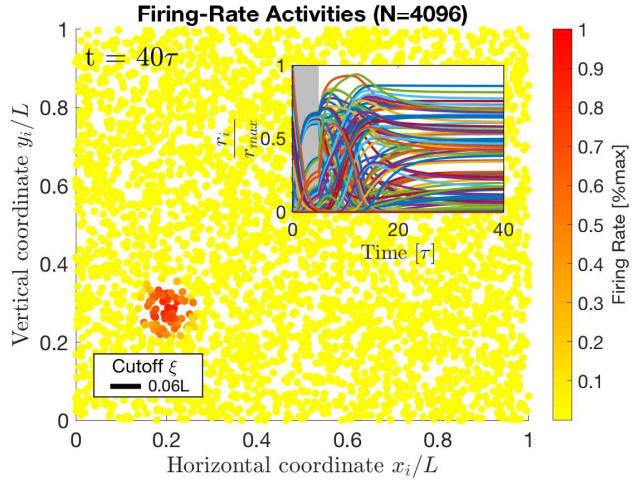}
  \caption{Sample bump state in a system with $N=2^{12}$. The scale bar indicates the synaptic cutoff distance $\xi$, below which $\mathcal{G}$ appears fully connected. ~\emph{Inset}: All the neural activities through time. Most of the trajectories remain near zero, and cannot be visually distinguished. Stimulation is shown as a gray block of width $\Delta t = 5\tau$.}
  \label{fig:visualization}
\end{figure}

Upon stimulation, the system initialized as above tends to develop a localized excitation in the vicinity of $\vec{x}_{\text{stim}}$, which quickly coalesces into a roughly circular ``bump" of activity~\cite{amari1977dynamics,compte2000synaptic,wang2001synaptic}. Figure~\ref{fig:visualization} depicts a representative bump in a system of size $N=2^{12}$ at $T=40\tau$. The inset reproduces the firing-rate trajectories for $t\leq T$, showing that all rates have stabilized to their final values by $T$.

While it is free to migrate or spread about $\vec{x}_{\text{stim}}$ during and after stimulation, this activity bump typically assumes a stable shape and location on the plane by the same time $T$. Analogous behaviors are observed when the system is stimulated from within a previously activated stable state. Then, activities associated with any preexisting bump are rapidly attenuated due to the global inhibition, typically returning to baseline activity values by $\Delta t$. Generally, given a sufficiently strong input current amplitude $A$ and adequately long stimulation time $\Delta t$, an activity bump will form in any general region of the plane and remain thereafter in the vicinity of $\vec{x}_{\text{stim}}$.

In simulation, our model seems to support only one spatially localized excitation under steady-state conditions, even if stimulated briefly at two locations simultaneously. At least qualitatively, this might be understood by analogy with a simpler system consisting of just two units, representing distant regions of strong firing. If each unit acts according to Eqs.~(\ref{eq:dynamics}-\ref{eq:activation}) -- loosely, as a self-excitatory, positive-feedback system, with a global inhibition that enters via the normalization $h_i / \sum_j h_j$ -- it is easy to imagine that their mutual feedback will lead to a single unit dominating (we ignore oscillations, since the feedback would need to be precisely tuned in order for these to appear). While it is not immediately clear from these equations that simultaneous activation at many locations will not lead inevitably to delocalized excitations or multiple small bumps, we are not focused on this here, precisely because we are interested in situations for which there is exactly one driving input at any given moment in time -- and only one recent memory, as in the experimental system of Ref.~\cite{kim2017ring}. Thus, as a rule of thumb, we say that the system supports a single bump at any given time~\cite{kim2017ring}, in any general spatial region of the plane.

How large are these activity bumps? Although they are not perfectly circular, we observe that excitations do take on a typical size for a fixed cutoff distance $\xi$. We can therefore speak about an effective bump radius $R_{\text{eff}}$. A simple way to measure $R_{\text{eff}}$ would be to choose a firing-rate threshold above which neurons will be considered \emph{active}, and compute the radius for the equivalent circular area $\pi R_{\text{eff}}^2$ occupied by this subset of system elements on the plane. Ideally, though, we would like to choose a criterion that is relatively insensitive to the cutoff distance. Fitting two-dimensional Gaussian curves to the spatial firing-rate distributions associated with each bump and measuring $2R_{\text{eff}}$ as the full width at half maximum, as done recently for the experimental system of Ref.~\cite{kim2017ring}, yields $R_{\text{eff}}(\xi=0.06L) \approx 0.78 \xi \approx 0.05 L$. In other words, the bump radius is on the order of the cutoff distance. We expect this to be a generic result.

Taking the ratio $\frac{R_{\text{eff}}}{\lambda} \approx 2.99$, we see that typical activity bumps are also large in comparison with the inter-neuron separation $\lambda$, as well as the distance $dL=10^{-2}L$ between adjacent gridpoints. This has an important consequence. If the system is stimulated at a point within (or too near) the area associated with an active bump, it may revert to the originally active bump state instead of evoking a new memory. This is particularly true if either the input time $\Delta t$ or amplitude $A$ are insufficiently large, but can occur more generally due to the fact that our random connectivity matrix lacks precise translational symmetry. This allows certain bumps to emerge as preferred states, which are more strongly favored than others (this limits the network representational capacity, as we determine quantitatively later). Nevertheless, the system does appear to select from a discrete, finite set of constant firing-rate states for the parameter values $(\lambda=N^{-\frac{1}{2}}$, $\xi \approx 3.84\lambda)$ defined above.

\vspace{0.1in}
In summary, for sufficiently strong input, we observe:
\begin{enumerate}
  \item Local stimulation can cause the system to develop stable bumps in essentially any region of the plane;
  \item The system seems able to transition, smoothly and repeatably, from sustaining one bump state to another (switch between multistable firing patterns);
  \item Independent stimulations centered at different gridpoints can result in nearly indistinguishable memory bumps.
\end{enumerate}

We take these observations together as the earmarks of dynamical attracting behavior -- in particular, the system acts as a discrete approximation to a 2D plane attractor. We identify each achievable bump state with a stored, retrievable memory. By definition, an attracting state persists until stimulation evokes a new bump, so we say that the system stores \emph{spatial memories} encoding the location at which it was most recently stimulated.

Since the basins of attraction (from within which stimulation at different $\vec{x}_{\text{stim}}$ values consistently leads to the activation of specific memories) are not infinitely small but instead appear finite, the system cannot remember arbitrary positions on the plane. It is then natural to ask how many \emph{unique} spatial locations can be distinguished by a given realization of the synaptic structure. That is, the resolution with which $\vec{x}_{\text{stim}}$ can be decoded requires quantification.

\subsection{Spatial memories span the entire plane}
\label{results__capacity}

How many distinct stimulation locations $\vec{x}_{\text{stim}}$ might we anticipate a realization $\mathcal{G}$ to resolve? We expect this \emph{capacity} to depend largely on gross statistics like the average size of the attracting basins, rather than on details of the instantial arrangement of neuron positions and synaptic connections associated with a given system configuration. 

Since the dynamical equations~(\ref{eq:dynamics}-\ref{eq:activation}) are deterministic, the attracting state evoked by stimulation at a given site should be unique, apart from the aforementioned dependencies on the initial state and input-current parameters. This variation can even be minimized: the stronger the external inputs, the more reliably we can anticipate that the system will find an attractor in the vicinity of the stimulation location, independent of where it is currently excited. Thus all that remains to determine the exact set of attractors supported by a given configuration $\mathcal{G}$ are the the coupling strengths. Accordingly, we expect that the bumps to which excitations attract will be almost exclusively a function of the (quenched) random variable $J_{ij}$.

We coarsely estimate the system's capacity as follows. Assuming homogeneous basins of attraction and one-to-one retrieval within a basin, the number of reliably stored memories will be equal to the number of basins that fit on the plane. Dividing the $L \times L$ space into equally-sized square sections of width $2R_{\text{eff}}^{-2}$ implies, for our parameter values, $\sim10^2$ distinct, nonoverlapping basins that span the 2D space. Thus our baseline will be $\sim 100$ bumps, touching tangentially.

A preliminary step towards more accurately quantifying the number of stimulation locations that the system can reliably encode is simply enumerating all the unique attractors activated during a given series of $n_{\text{trials}}$ stimulations. This allows us to conceptualize the capacity in terms of input (stimulation site) to output (bump location) relations. For each stimulation, we track the \emph{center of excitation} $\vec{x}_{\text{COE}}(t)=\sum_{i^\prime\frac{r_{i^\prime}(t)\vec{x}_{i^\prime}}{aN}}$ among cells $i^\prime$ which we identify as actively participating. Instead of accommodating for the uncertainties associated with Gaussian fits, here we employ simple thresholding to identify active units, for two principal reasons. First, even the fixed-threshold criterion $r_i>10a$ predicts the number of active neurons to within $10$ units of the amount given by the \emph{participation number} $p_\nu=(\sum_{i=1}^N r_i^\nu)^2/\sum_{i=1}^N r_i^{2\nu}$, and it exhibits similar qualitative behavior across the surprisingly large range of cutoffs from roughly zero to $10\lambda$. In addition, this criterion was found to predict coordinates for the excitations that coincide well with the measured Gaussian peaks.

For large cutoffs, it is possible that even a fairly nonrestrictive threshold can exclude relatively strongly firing neurons: our constraint $\sum_i r_i(t)=a$ implies that firing activity within a given bump decreases as bumps increase in size, which is precisely what we observed to happen as we increase $\xi$. Excitations encompassing zero active neurons were to be assigned a special value of $\vec{x}_{\text{COE}}(t)$, allowing us to count them separately toward the capacity, but this was not observed for the $\xi=0.06L$ presented below. We enumerate all distinct bumps by counting the unique values of $\vec{x}_{\text{COE}}(T)$ observed, to within a specific resolution (we discuss the importance of this resolution below). For $n_{\text{trials}}$ large, this number should approach the cardinality of the set of possible memories. The next step will be to quantify how many -- or with what fidelity -- distinct values of the gridpoint coordinates $\vec{x}_{\text{stim}}$ can be discriminated by these enumerated attractors.

We measure the capacity for a given realization $\mathcal{G}$ as follows. Although each site in the set of $(L\cdot dL)^{-2} = 10^4$ available stimulation gridpoints is visited $\frac{n_{\text{trials}}}{L\cdot dL^{-2}}=k$ times each in each series of stimulation events, averaging over all possible initial conditions for each gridpoint would require too much time. Here we choose $k=10$ to further mitigate finite-sampling errors due to the situation described above, in which stimulation near a highly active bump simply reverts the system back to that previous attractor after a transient. We also choose to work with an information-theoretic capacity metric, to treat the inherently nonuniform stochasticity associated with the ``stimulus-response" records in a natural framework.

Specifically, we measure the mutual information~\cite{shannon1948mathematical} between random variables $\vec{x}_{\text{stim}}$ and $\vec{x}_{\text{COE}}(T)$ for a realization $\mathcal{G}$. To do this, we obtain the frequencies of occurrence for all observed stimulation locations $\lbrace \vec{x}_{\text{stim}} \rbrace$ and bump centers $\lbrace \vec{x}_{\text{COE}}(T) \rbrace$, over a set of $n_{\text{trials}}$ stimulation events. We then use these frequencies as the maximum-likelihood estimates of the corresponding probabilities to form the ``plug-in" or na\"ive estimators for the relevant entropies~\cite{antos2001convergence,strong1998entropy,paninski2003estimation}, from which we can calculate the mutual information $MI\left(\lbrace\vec{x}_{\text{stim}}\rbrace,\lbrace \vec{x}_{\text{COE}}\right)(T)\rbrace$. Since asking how many different attractors were observed for each stimulation position is equivalent to asking how many different stimulation positions lead to the same attractor (i.e., the mutual information is symmetric), we choose the latter. Finally, from the mutual information, we define the capacity
\begin{equation}
    C = 2^{MI\left(\lbrace\vec{x}_{\text{stim}}\rbrace,\lbrace \vec{x}_{\text{COE}}(T)\rbrace\right)}.
\end{equation}

\begin{figure}[ht]
  \centering
  \includegraphics[width=1\linewidth]{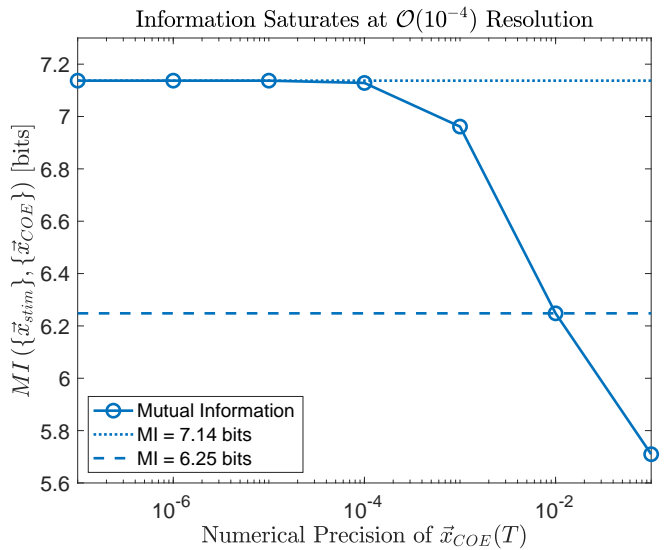}
  \caption{Mutual information as a function of rounding precision in the center-of-excitation values $\lbrace \vec{x}_{\text{COE}}(T) \rbrace$. Saturation occurs by four decimal places, but in what follows we keep two places to ensure the precision of $\vec{x}_{\text{COE}}$ is not finer than the inter-neuron separation $\lambda$. The changes the capacity by less than a factor of 2.}
  \label{fig:precision}
\end{figure}

Since the information is measured over discrete states, we must discretize the the values of $\vec{x}_{\text{COE}}(T)$ by rounding them to an appropriate resolution. As seen in Fig.~\ref{fig:precision}, truncating $\vec{x}_{\text{COE}}(T)$ to two decimal places still represents $87.5\%$ of the maximum information, or $\approx 6.25$ bits. Assuming that the system cannot track bump centers to a precision better than these two decimal places -- roughly the theoretical separation between neurons -- we arrive at $C \approx 76$ distinct stimulation regions for the values of $L$, $\lambda$, $\xi$ and $\rho$ used throughout.

\begin{figure}[ht]
  \centering
  \includegraphics[width=1\linewidth]{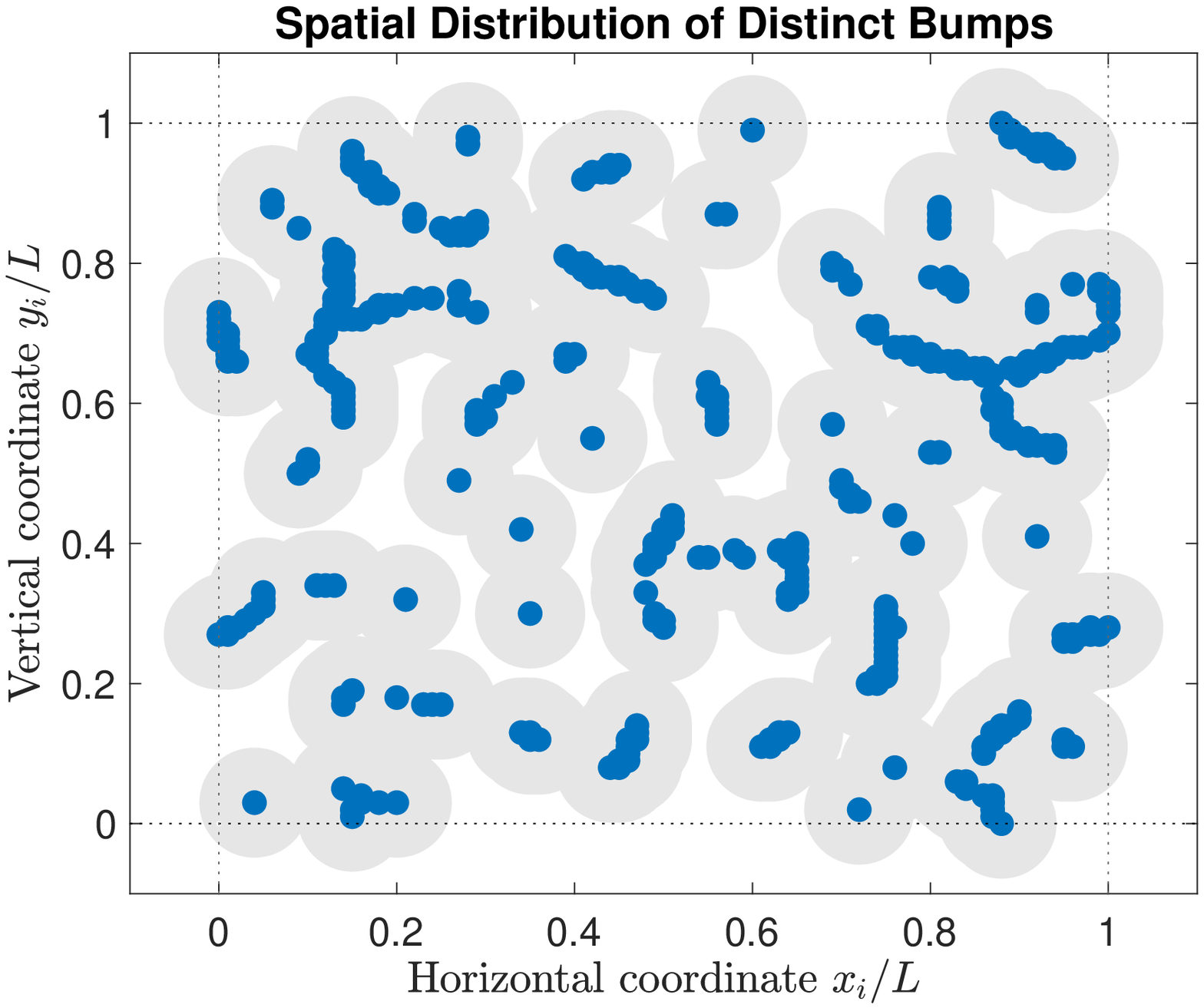}
  \caption{The different attracting bumps observed over the $k \cdot n_{\text{trials}}$ computational experiments are distributed in such a way that they span the majority of the plane. Bump centers are shown as blue dots; radii for their surrounding gray circles are $\approx R_{\text{eff}}$. Dotted lines are periodic boundaries.}
  \label{fig:whitespace}
\end{figure}

In other words, on average, $\mathcal{G}$ is able to store and reliably retrieve a number of memories approximately equal to our na\"ive, baseline estimate. Unlike in that coarse estimation, we did not require bumps to be nonoverlapping in measuring the capacity -- yet the system's recall ability turns out to be nearly as accurate as a fully deterministic discriminator that simply decides in which $R_{\text{eff}} \times R_{\text{eff}}$-sized, homogeneous division of the plane the last stimulation occurred. Thus the information-theoretic capacity, measured to two decimal digits precision in $\vec{x}_{\text{COE}}(T)$, is also consistent with a typical size for the attracting basins which matches $R_{\text{eff}}$ for stable bumps. Furthermore, we observe that the retrievable memories span more or less the entire spatial extent of the $L \times L$ plane. This can be readily observed in Fig.~\ref{fig:whitespace}, which depicts the set $\lbrace \vec{x}_{\text{COE}} \rbrace$ of unique bumps accounted for over a course of $n_{\text{trials}}$ stimulations for one network realization.

\subsection{Mutual information is near-optimal for a broad range of parameter values}

The cutoff distance is an important length scale in the system. The structure of the network depends crucially on $\xi$, allowing us to go from completely unconnected neurons in the extreme of $\xi=0$ to the fully-connected network for $\xi = L$. It is important to understand how $\xi$ affects our main findings -- in particular, the existence of localized excitations, and the number of memories $\mathcal{G}$ can support.

For the unconnected case $\xi = 0$, we have $\lbrace J_{ij} \rbrace =0$. In the absence of recurrent connections (besides the implicit inhibition), all neurons respond independently to their respective external inputs $I_i(t)$: that is, the $\lbrace{r_i\rbrace}$ obey a simplified version of Eqs.~(\ref{eq:dynamics}-\ref{eq:activation}). In order to write down the dynamics in this case, we first note that neurons outside the stimulation patch have activations $h_i = f(0) \approx 0$ for both $t<\Delta t$ and $t\geq \Delta t$. These units at first experience an exponential decay in their firing activities and then approach the steady-state value $r_i(t\gg \Delta t)=a$. The $\sim \pi N (\rho/L)^2$ neurons encompassed by the stimulation patch also approach a constant value. To show this, we note that each of the units in this latter subset sees the same input $h_i = f\left[ A \left( 1- \Theta(t-\Delta t) \right) \right]$, so that the ratio $h_i(t)/\sum_j h_j(t)$
stays constant. Therefore we can remove the nonlinearities entirely and write
\begin{align}
\frac{dr_i}{dt} &= -r_i + I_i^\prime(t), \label{simplified_dynamics}
\\
I_i^\prime(t) &=  \left\{
        \begin{array}{ll}
           \frac{a}{\pi \rho^2}, & t < \Delta t, \\
            a, & t \geq \Delta t.
            \label{eq:newinput}
        \end{array}\right.
\end{align}
Then, in the long-time limit, the unconnected system relaxes to the trivial stable state $\lbrace r_i(t \gg \tau) \rbrace = a$, in which all neurons fire at the same, baseline rate. It cannot sustain any excitations that can be decoded as memories. In the other extreme, $\xi \rightarrow L$, it seems unlikely that a fully-connected network can support  any \emph{localized} excitations.

\begin{figure}[ht]
  \centering
  \includegraphics[width=1\linewidth]{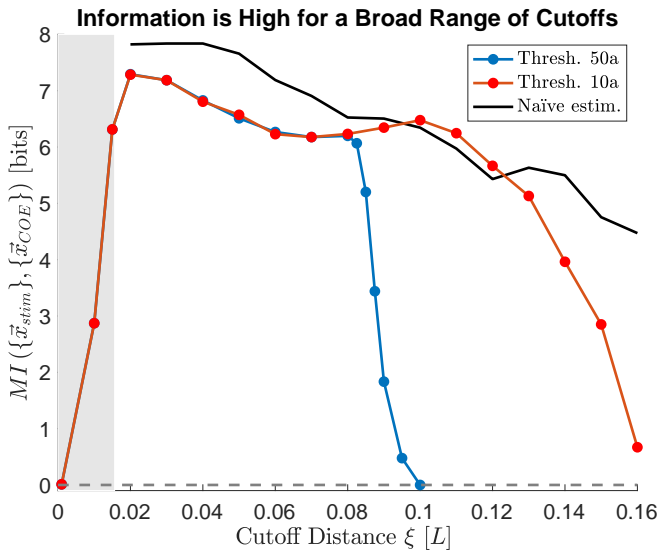}
  \caption{Varying $\xi$ reveals a broad plateau over which the mutual information remains within a single bit of its maximum value. At either extreme of $\xi$ the information falls to zero as connectivities become too sparse or too dense to support the type of spatial memory discussed throughout. The black curve represents the information $\log_2{\frac{L^2}{\pi R_{\text{eff}}}}$ corresponding to our original, na\"ive estimate of $C$, with $R_{\text{eff}}(\xi)$ adjusted to match the typical values given by Gaussian fits to $\sim1000$ bumps. Note that the black curve, representing $\xi<\lambda$, exists only outside the shaded gray box because the bumps that did localize for small $\xi$ were too few to measure $R_{\text{eff}}(\xi)$ accurately.}
  \label{fig:cutoff_transition}
\end{figure}

We quantify the precise dependence of our  findings on the value of the cutoff distance in Fig.~\ref{fig:cutoff_transition}. We generated this plot by progressively decreasing $\xi$ for an initial, fully-connected realization $\mathcal{G}$. Here we chose $k=1$, stimulating at the first $10^4$ of the $10^5$ sites used to generate Fig.~\ref{fig:precision}, and rounded the measured information values to a precision of two decimal places in $\vec{x}_{\text{COE}}$ as decided above. Clearly, the mutual information quickly drops to zero below the inter-neuron separation $\lambda$. This means that the system attains only states that are delocalized -- effectively all neurons contribute to the excitation, but none exceed the threshold $r_i>10a$ to be considered ``active" -- which we identify as the single, trivial state.

At the other extreme, the mutual information returns to zero for large values of $\xi$. This can be explained in terms of the circumstances discussed in Section~\ref{results__supportsattractors}, in which it becomes difficult for the network to switch out of its preferred states. As the cutoff distance increases above $\xi\approx 7\lambda$ (or $\approx 5\lambda$ for the stricter threshold of $r_i=50a$), more neurons are directly involved in sustaining a given excitation, and the structure of the basins of attractions changes so as to accommodate fewer feasible memories. As in the case of insufficient stimulation time or amplitude, the success or failure of a given stimulation in evoking a nearby bump is somewhat history-dependent (in the sense that some memories might be retrievable from some initial states but not certain preferred states), but invariably the system comes to favor a single state in the limit that the network becomes fully connected. For the $10a$ threshold, the network cannot reliably store any spatial memories for roughly $\xi>0.16L \approx 10\lambda$.

Between these two extremes, there is an optimal value $\xi^* \approx 0.02L$, for which the greatest number of stimulation gridpoints can be distinguished. Moreover, starting at this value, there is a plateau in the system's accuracy from roughly $\xi = 0.02L \dots 0.11L \approx \lambda \dots 7 \lambda $, across which the mutual information varies by only $\sim 1$ bit. More precisely, the gap between the highest and lowest points on the $10a$-threshold curve of Fig.~\ref{fig:cutoff_transition} corresponds to the difference between resolving $C\approx 156$ and $C \approx 72$ distinct stimulation sites. These values are of the same rough order of magnitude, and their average is nearly equal to our very first baseline estimate of $100$ distinct, homogeneous basins. We note in particular that the cutoff distance $\xi=0.06L$ used throughout the rest of the paper is nominally three times larger than $\xi^*$, but different by less than the aforementioned bit in terms of information.

\begin{figure}[ht]
  \centering
  \includegraphics[width=1\linewidth]{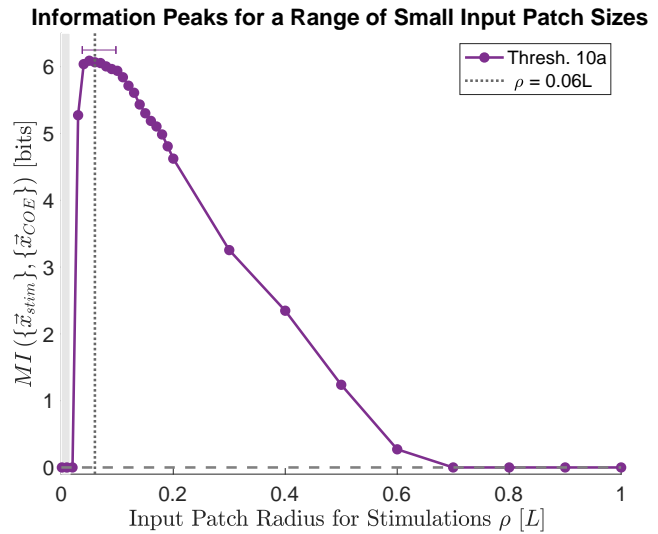}
  \caption{In the neighborhood of $\rho = \xi = 0.06L$, the mutual information does not vary significantly. We verified that the system tends to fall into the same attractor regardless of the specific value of $\rho$ until a large percentage of neurons are stimulated, thereby activating the aforementioned ``preferred" or global states. At roughly the same value after which see a decrease in information with the cutoff distance, we observe a drop in information with $\rho$. This continues monotonically until $\rho > 50\lambda$, after which stimulations leads only to excitations below the activity threshold.}
  \label{fig:rho}
\end{figure}

In principle, the capacity should also depend on how reliably the system accesses its attractors for (or indeed, whether the set of accessible attractors changes with) different values of the size of the input patch, $\rho$. Figure~\ref{fig:rho} records the dependence of the mutual information on $\rho$. Outside this range, the system will attract to (possibly different) preferred states, but between roughly $2\lambda$ and $6\lambda$ we observe that the system attracts to the same bump state regardless of the specific value of $\rho$ (not explicitly depicted). This gives the appearance that the system really is tracking the stimulation centers in computing its final states, at least for input patch sizes in this range.

To the extent that different proxies for $\vec{x}_{\text{COE}}$ agree, this suggests that the system does in fact encode a coarse representation of the stimulation location -- the bump centers of excitations -- rather than tracking high-dimensional quantities like the real-valued firing rates. That is, although an experimental system wired according to our prescription for $\lbrace J_{ij} \rbrace$ could indeed store information in individual firing rates for other purposes, we are not merely imposing but discovering that the low-dimensional summary variable $\vec{x}_{\text{COE}}$ is sufficient to predict the stimulation region to a considerable accuracy. Another step toward testing this hypothesis would be to systematically map the basins of attraction for a given realization $\mathcal{G}$, and check whether the steep decrease shown in Fig.~\ref{fig:rho} occurs when the stimulation patch grows large enough to extend into multiple basins besides that of the targeted memory.

Together, the above results suggest that our randomly-weighted network can sustain local excitations for a range of parameter values. In general, these excitations can serve reliably as spatial memories encoding the system's most recent stimulation location if the number of neurons activated via stimulation and local synaptic input is small relative to the system size $N$. This can be achieved by choosing $\xi$ less than approximately $\mathcal{O}(10\lambda)$, which ensures that a given neuron synapses with anywhere from roughly $\pi(\lambda)^2\sigma$ \dots $\pi(10\lambda)^2\sigma \approx 10^0 \dots 10^2$ neighbors.

\section{Discussion}
\label{discussion}

We have showed that short-range, but otherwise unstructured connectivities can support spatial memory via persistent firing if the overall activity of the network is constrained through excitation-inhibition balance. The spatial regions that can be remembered (discriminated) with high-fidelity effectively tile our $L \times L$ planar section, with a resolution of $\mathcal{O}(\lambda^{-1})$ distinct sites, roughly equivalent to the number of nonoverlapping memories that span the same area. This performance corresponds to an information-theoretic capacity that scales as $C \propto \sqrt{N}/L$, or $C \propto \sqrt{\sigma}$ in terms of the neuron density, which can be checked experimentally by testing larger system sizes.

Since the inter-neuron separation sets the scale of the problem at the outset, it is not necessarily surprising that the optimal cutoff distance $\xi^* \approx \lambda$. What is unexpected in our results is the fact that a spatial memory spanning a two-dimensional manifold can be achieved without explicit tuning of synaptic connections. This is reinforced by the fact that we observe not just an isolated peak at $\xi^*$, but a broad plateau of near-optimal cutoff distances.

While it is traditionally maintained~\cite{zylberberg2017mechanisms} that only tuned connectivity profiles can produce continuous attractors, the idea that random networks support memory on short time scales is not altogether new~\cite{griffith1963stability,maass2002real,buonomano2009state,kriener2014dynamics}. Indeed, recent work argues that quasi-random topologies, refined via a non-linear Hebbian learning rule, can give rise to attractor dynamics in the specific context of persistent neural activity as a substrate for working memory~\cite{pereira2018attractor}. However, none of these examples attempt to store spatial memories that effectively span a continuous manifold. In addition, we accomplish spatial memory using a random network, which emphatically requires no learning.

Similarly, distance-dependent topologies~\cite{gerstner2014neuronal} have been implemented in previous models, including the seminal work on continuous neural attractors~\cite{amari1977dynamics}, yet we are aware of only two related studies that link sparse, short-range (1D nearest-neighbor) connections formally to the localization of firing-rate excitations~\cite{amir2016non,tanaka2019non}. As we do, both respect Dale's Principle~\cite{catsigeras2013dale} for the signs of synaptic connections only indirectly~\cite{zhu2015modeling} and explore random weights. While it may be interesting to explore the spectra of our $\lbrace J_{ij} \rbrace$ in the context of Anderson localization or the notion of ``spatially structured" disorder developed in~\cite{tanaka2019non}, a more obvious generalization of our model would be to relax the hard-threshold cutoff condition to a connection probability. For example, we could set $J_{ij} \propto e^{-|\vec{x}_i - \vec{x}_j|/\xi}$, or another function of $d_{ij}=\lVert\vec{x}_i - \vec{x}_j\rVert$.

A drawback to our model, in the form presented here, is that the system of Eqs.~(\ref{eq:J}-\ref{eq:input}) incorporates no explicit noise terms. Fundamental to our results is the firing-rate constraint $\sum_i r_i(t)= aN$, an imposition which corresponds only approximately to the biological reality for real circuits (as in~\cite{kim2017ring}). In our future work, we propose to replace the constant parameter $a$ by a Gaussian process $\alpha(t)=a+\eta(t)$. We expect that, for small amounts of noise, the system will retain its qualitative behavior, but with a reduced capacity. On the other hand, for $\eta(t)$ with large variance, it is possible that the system will fail to store memories with high fidelity due to longer bump excursions or delocalization, or entirely as with $\xi$ and $\rho$.

If these assumptions regarding the inclusion of noise are found to hold, it would be interesting to explore noise parameters that place the firing-rate variability in a regime consistent with previous experiments~\cite{barbieri2008can,zylberberg2017mechanisms} while respecting our sparsity constraints. Yet we reiterate that our goal is not to model any known experimental system. Indeed, whether or not our model relates to specific, observable experimental systems remains to be seen. In anticipation of such \emph{in vivo} analogs, we offer the following predictions regarding which features of our model might be used to infer whether short-range, randomly weighted connections drive a given instance of persistent activity.

First, in the best case scenario, novel technologies may allow researchers to probe structural properties directly. This promises a trivial way of checking whether synaptic matrices are untuned, as in Eq.~(\ref{eq:J}), and is already underway for the fly~\cite{turner2017angular,franconville2018building,li2019automated}. While the emerging picture for \emph{Drosophila} is one of decidedly nonrandom connectivity, this may not hold for significantly larger organisms. Indeed, the number of possible synapses in a neural system scales as $\mathcal{O}(N^2)$. Thus genetic encoding of precise values for some billions of pairwise connections even in modestly sized vertebrates is simply not feasible. On the other hand, it is plausible that regularity appears at the level of local rules superimposed on essentially random connectivities, as in canonical microcircuit models~\cite{sporns2011non}, which would be consistent with our setup.

In the absence of structural information, the firing-rate activities themselves can also help 
support or reject our model. Since most classic continuous-attractor architectures have translationally invariant connections, they are able to host bumps at virtually any location~\cite{wu2016continuous}.
Our $\lbrace J_{ij}\rbrace$, on the other hand, lack such a symmetry. This leads to discrete attractors~\cite{inagaki2019discrete} with variable spacing and portions of the plane that cannot be reliably encoded. Such ``discrete approximations" to attracting manifolds have even been touted as more robust than their continuous counterparts, for example to perturbations in the synaptic weights~\cite{kilpatrick2013optimizing}. It would be interesting to quantify the fraction or extent of the plane that the system can remember in the presence of the aforementioned noise.

In addition, while continuous attractor models accommodate a degree of drift or diffusion for activity bumps following their settlement upon the manifold~\cite{wimmer2014bump}, tracking $\vec{x}_{\text{COE}}(t)$ reveals that excursions in our random networks occur predominantly \emph{before} $t=T$; see the inset of Fig.~\ref{fig:visualization}. Thus, comparing the observed distribution of displacements, between the tested $\vec{x}_{\text{stim}}$ values and the corresponding $\vec{x}_{\text{COE}}(t)$ could also distinguish our model.

Finally, the raw activity measurements $\lbrace r_i(t) \rbrace$ are also subject to what is known as network reverse-engineering, or automated inference methods that operate directly on data to reconstruct network interaction structures~\cite{natale2018}. Although we do not advocate applying out-of-the-box algorithms to glean structural information in general, there do exist certain signatures and gross statistics which can be used to differentiate truly random graphs from more complex or subtle architectures at a coarse level~\cite{estrada2011}.

Our model is one of many that attempt to capture the ability of different neural systems to support localized excitations that encode real-valued quantities. Here, we eschew structured topographic mappings~\cite{kim2017ring} in favor of a random connectivity that we find to be capable of storing similar neural representations. Whether or not \emph{in vivo} circuits conforming to the specifications of our model are found experimentally to underlie one of these interesting systems, in our view such random, balanced excitatory-inhibitory networks should still be taken seriously as null models for recurrent neural computation~\cite{sederberg2019randomly}.

\vspace{0.2in}
\begin{acknowledgments}

This work was supported in part by the following grants: JSMF/220020321 and NSF/IOS/1208126, NSF/BCS/1822677, and NIH 1 R01 EB022872.

\end{acknowledgments}

\bibliography{bibliography}

\end{document}